\documentclass[9pt]{article}
\usepackage{amsfonts}
\usepackage{pstricks}
\usepackage{pst-node}
\usepackage{epsfig}
\usepackage{epsfig}
\usepackage[latin5]{inputenc}
\usepackage[T1]{fontenc}
\usepackage{lipsum}

\begin{document}
	\def\ba{\begin{eqnarray}}
	\def\ea{\end{eqnarray}}
	\def\w{\wedge}
	\def\d{\mbox{d}}
	\def\D{\mbox{D}}

\begin{titlepage}	
\title{A non-minimally coupled, conformally extended Einstein-Maxwell theory of  pp-waves}
\author{Tekin Dereli\footnote{tdereli@ku.edu.tr},     
Yorgo \c{S}eniko\u{g}lu\footnote{ysenikoglu@ku.edu.tr}
\\
{\small  Department of Physics, Ko\c{c} University, 34450 Sar{\i}yer-\.{I}stanbul, Turkey}  }
 \maketitle	

	

\begin{abstract}
\noindent
A non-minimal coupling of Weyl curvatures to electromagnetic fields is considered in Brans-Dicke-Maxwell theory. The gravitational field equations are formulated in a Riemannian spacetime where the spacetime torsion is constrained to zero by the method of Lagrange multipliers in the language of exterior differential forms. The significance and ramifications of non-minimal couplings to gravity are examined in a pp-wave spacetime. 
 \end{abstract}



\end{titlepage}

\newpage

\section{Introduction}
Einstein's theory of general relativity is a theory of gravitation determined by the Riemannian geometry of a 4-dimensional spacetime. Brans-Dicke theory \cite{B1,B2,B3} on the other hand, provides a modification through the inclusion of a massless scalar field, in order to incorporate Mach's principle into Einstein's initial framework. The coupled field equations of the theory may be obtained by field variational principle from an action. It is well known that the second order principle where the metric variations of the Levi-Civita connections are taken into account; and the first order (Palatini) where independent variations of the action relative to the metric and connection yield, in these cases, the same set of field equations \cite{B8}.The fact that the connection is Levi-Civita may be imposed by the method of Lagrange multipliers.
\vspace{2mm}

\noindent
Electromagnetic fields can be coupled minimally to gravitation simply by incorporating the Maxwell Lagrangian density 4-form into the total action; and taking independent variations of the electromagnetic fields. Here we consider non-minimal couplings of gravity and electromagnetic fields in the Lagrangian density. Such non-minimal considerations for the Einstein-Maxwell couplings may be justified in extreme conditions where there are intense gravitational and electromagnetic fields at high temperatures, pressures and density \cite{B7-a,B7-b,B7-c,B7-d}. 

\vspace{2mm}
\noindent
We will use in what follows a first order Palatini type variational principle where the Levi-Civita connection is forced by zero-torsion constraint using the method of Lagrange multipliers \cite{B9-a,B9-b}. In a similar way, we may avoid the use of an electromagnetic potential 1-form $A$ such that $F=dA$ by imposing the closedness of $F$, $dF=0$, by the same method and varying $F$ directly. We have previously discussed \cite{B10}(work in progress) in the non-minimal coupling context, static, spherically symmetric field configurations where electromagnetic charges will be screened by gravitational fields\cite{B10b}. We have also investigated pp-wave spacetimes in Brinkmann coordinates \cite{B11a} for one simple case of $RF^2$ couplings which induced possible modifications to the metric functions\cite{B16}. Here we consider non-minimal couplings of $RF^2$ type to the Brans-Dicke scalar-tensor theory of gravity. 

\vspace{2mm}
\noindent
The Brans-Dicke-Maxwell Lagrangian  
\ba
\mathcal{L}_0&=&\frac{\phi}{2}R_{ab} \wedge *(e^a \wedge e^b) -\frac{\omega}{2\phi} d\phi \wedge *d\phi - \frac{1}{2}F \w *F.
\ea
is locally scale invariant under 
\ba
g \rightarrow e^{2\sigma}, \quad \phi \rightarrow e^{-2\sigma} \phi, \quad F \rightarrow F,
\ea
in the special case $\omega=-\frac{3}{2}$.
The non-minimal $RF^2$ coupling of curvature to electromagnetic fields that will be considered here is
\ba
\mathcal{L}_1&=&\frac{\gamma}{2\phi}C_{ab}\w F^{ab}*F
\ea
where 
\ba
C_{ab}=R_{ab}-\frac{1}{2}(e_a\w R_b - e_b \w R_a) + \frac{R}{6}e_{ab}
\ea
are the Weyl conformal curvature 2-forms and $\gamma$ is an arbitrary dimensionless coupling constant. $\mathcal{L}_1$ is also locally scale invariant.
\section{Field Equations}
In a 4-dimensional spacetime manifold, the field equations will be derived by infinitesimal variations of the action $I[e^a,\omega^{a}_{\;\;b},\phi,F,\lambda_a,\mu]=\int_{M} (\mathcal{L}+\mathcal{L_{C}}),$ where the Lagrangian density 4-form
\ba
\mathcal{L}&=&\frac{\phi}{2}R_{ab} \wedge *(e^a \wedge e^b) -\frac{\omega}{2\phi} d\phi \wedge *d\phi - \frac{1}{2}F \w *F \nonumber\\ 
&+& \frac{c_1}{2\phi}R_{ab}F^{ab} \w *F +\frac{c_2}{2\phi} F_a \w R^a \w *F + \frac{c_3}{2\phi}RF \w *F.
\ea

\noindent $\{e^a\}$ 's are the co-frame 1-forms in terms of which the space-time metric $g=\eta_{ab}e^{a} \otimes e^b$ with $\eta_{ab}=diag(-+++)$. ${}^*$ denotes the Hodge map. The orientation is given by $*1 = e^0 \wedge e^1 \wedge e^2 \wedge e^3$. We denote the Brans-Dicke scalar field by $\phi$ with its parameter $\omega$ and the electromagnetic field 2-form by $F$. The non-minimal dimensionless coupling constants are $c_1,c_2$ and $c_3$. $\{{\omega}^{a}_{\;\;b}\}$ are the connection 1-forms that satisfy the Cartan structure equations
 \ba
 de^a + \omega^{a}_{\;\;b}  \wedge e^b = T^a
 \ea
 with the torsion 2-forms $T^a$ and  
 \ba
 d \omega^{a}_{\;\;b} + \omega^{a}_{\;\;c}  \wedge \omega^{c}_{\;\;b} =R^{a}_{\;\;b}   
 \ea
with the curvature 2-forms $R^{a}_{\;\;b}$ of spacetime.
The field equations are retrieved by independent variations of the action with respect to $e^a$, ${\omega}^{a}_{\;\;b}$, $\phi$ and $F$. In order to implement the method of Lagrange multipliers, we add to the Lagrangian density above the constraint terms
\ba
{\mathcal{L_C}}=( de^a + \omega^{a}_{\;\;b}  \wedge e^b) \w \lambda_a + dF \w \mu.
\ea
We impose the connection to be, through constrained variations, the unique metric compatible, torsion-free Levi-Civita connection. Constrained by the Lagrange multiplier 2-form $\mu$, the homogeneous Maxwell equations $dF=0$ are satisfied by the electromagnetic 2-form $F=\frac{1}{2}F_{ab} e^a \w e^b$.
The infinitesimal variations of the total Lagrangian density with respect to the co-frames $e^a$, the scalar field $\phi$, the electromagnetic field $F$ and the connection 1-forms ${\omega}^{a}_{\;\;b}$ give, respectively, the following equations
\newpage
\ba
&&\frac{\phi}{2}R^{bc}\w*e_{abc} + \frac{1}{2}(\iota_aF \w *F - F \w \iota_a *F) + \frac{\omega}{2\phi}(\iota_ad\phi \w *d\phi + d\phi \w \iota_a *d\phi) +D\lambda_a \nonumber \\
&&-\frac{c_1}{4\phi}\big[-4F_{ac}F_b \w *R^{cb}+\iota_aR_{bc}\w F^{bc}*F-R_{bc}F^{bc}\w \iota_a*F+F^{bc}F_a \w *R_{bc}-F^{bc}F \w \iota_a*R_{bc}\big] \nonumber\\
&&+\frac{c_2}{4\phi}\big[-F_{ac}R^c\w*F+2RF_a\w*F-\iota_aR^cF_c\w*F+2\iota_aR^{cb}F_{cb}\w*F-2\iota_aR^{cb}\w F_c \w \iota_b *F\nonumber \\
&&-F_a\w*(F_c\w R^c)-2F_a\w R^c \w \iota_c *F +F_c \w R^c \w \iota_a*F + F \w \iota_a * (F_c \w R^c)\big] \nonumber \\
&&+\frac{c_3}{2\phi}\big[-2\iota_aR^bF_b\w*F-2\iota_aR^bF\w \iota_b*F-RF_a\w *F+RF\w \iota_a *F\big]=0, \label{coframe-eq}
\ea
\ba
&&\frac{1}{2}R_{ab}\w *e^{ab} - \frac{1}{2\phi^2}(c_1R_{ab}F^{ab}\w*F + c_2F_a \w R^a \w *F + c_3 RF\w*F)\nonumber \\
&&+ \frac{\omega}{2\phi^2} d\phi \w *\d\phi + \omega d(*\frac{\d\phi}{\phi})=0,\label{phi-eq}
\ea
\ba
&&\frac{1}{\phi}\big[c_1F^{ab}*R_{ab}+\frac{c_2}{2}(R^a\w \iota_a *F - R*F + *(F_a\w R^a)) +c_3 R*F\big]-*F -d\mu=0,\nonumber\\ \label{em-eq}
\ea
\ba
&&\frac{1}{2}D\big[\phi *e_{ab}+c_1(\frac{F_{ab}*F}{\phi})+c_2(-\frac{F_{ab}*F}{\phi}+\frac{1}{2\phi}F_a\w \iota_b*F -\frac{1}{2\phi}F_b \w \iota_a*F) \nonumber \\
&&+\frac{c_3}{\phi}\iota_b\iota_a(F\w*F)\big]+\frac{1}{2}(e_b\w\lambda_a-e_a\w\lambda_b)=0.\label{con-eq}
\ea
D denotes the covariant exterior derivative with respect to the Levi-Civita connection and $\iota_a$ are the interior products that satisfies $\iota_a e^b=\delta_a^b$.
Variations of the Lagrange multiplier 2-forms $\lambda_a$ impose the zero-torsion condition $T^a=0$.

The field equation from the variation of the constraint 1-form $\mu$ gives the homogeneous Maxwell equations $dF=0$. Equivalently by the Poincaré Lemma, $F=dA$ for some unique potential $A$.
We read the electromagnetic field equation from (\ref{em-eq}) by taking its exterior derivative

\ba
d*F=d\Big[\frac{1}{\phi}\big(c_1F^{ab}*R_{ab}+\frac{c_2}{2}(R^a\w \iota_a *F - R*F + *(F_a\w R^a)) +c_3 R*F\big)\Big]. \label{mu-eq}
\ea

\vspace{4mm}
\noindent
Multiplying equation (\ref{phi-eq}) we obtained from the $\phi$ variations by $2\phi$ and subtracting it from the trace of the co-frame $e^a$ variation equation (\ref{coframe-eq}), we simplify the scalar field equation to the form
\ba
2\omega d(*d\phi)-e^a \w D\lambda_a=0.
\ea
 
\vspace{4mm}
\noindent
$\lambda_a$'s are Lagrange multiplier 2-forms to be solved from the connection variation equations (\ref{con-eq}) that is written as
\ba
(e_a\w\lambda_b-e_b\w\lambda_a) = d\phi \w*e_{ab} + K_{ab}
\ea
where
\ba
K_{ab}=D(\frac{1}{\phi}\Sigma_{ab})
\ea
and
\ba
\Sigma_{ab}=c_1F_{ab}*F+c_2(-F_{ab}*F+\frac{1}{2}F_a\w \iota_b*F -\frac{1}{2}F_b \w \iota_a*F)+c_3\iota_b\iota_a(F\w*F).
\ea
Then we obtain the following expression for the Lagrange multipliers
\ba
\lambda_a=\iota_a(*d\phi) + \iota_bK^b_{\;a} - \frac{e_a}{4}\w\iota_c\iota_bK^{bc}. 
\ea
The Einstein 3-forms are given by $G_a=-\frac{1}{2}R^{bc}\w*e_{abc}$, so that the Einstein field equations can be written as
\ba
&&\phi G_a= \nonumber \\
&&\frac{1}{2}(\iota_aF \w *F - F \w \iota_a *F) + \frac{\omega}{2\phi}(\iota_ad\phi \w *d\phi + d\phi \w \iota_a *d\phi) +D(\iota_a*d\phi) \nonumber\\ 
&&-\frac{c_1}{4\phi}\big[-4F_{ac}F_b \w *R^{cb}+\iota_aR_{bc}\w F^{bc}*F-R_{bc}F^{bc}\w \iota_a*F+F^{bc}F_a \w *R_{bc}-F^{bc}F \w \iota_a*R_{bc}\big] \nonumber\\
&&+\frac{c_2}{4\phi}\big[-F_{ac}R^c\w*F+2RF_a\w*F-\iota_aR^cF_c\w*F+2\iota_aR^{cb}F_{cb}\w*F-2\iota_aR^{cb}\w F_c \w \iota_b *F\nonumber \\
&&-F_a\w*(F_c\w R^c)-2F_a\w R^c \w \iota_c *F +F_c \w R^c \w \iota_a*F + F \w \iota_a * (F_c \w R^c)\big] \nonumber \\
&&+\frac{c_3}{2\phi}\big[-2\iota_aR^bF_b\w*F-2\iota_aR^bF\w \iota_b*F-RF_a\w *F+RF\w \iota_a *F\big] \nonumber \\ 
&&+ D(\iota_bK^b_{\;a}) + \frac{e_a}{4}\w D(\iota_c\iota_bK^{bc}).
\ea
We have to solve them together with the scalar field equation
\ba
(2\omega+3)d(*d\phi)=d(\iota_bK^b_{\;a}\w e^a). \label{sc-eq}
\ea
and the inhomogeneous Maxwell equations
\ba
d*F=d\Big[\frac{1}{\phi}\big(c_1F^{ab}*R_{ab}+\frac{c_2}{2}(R^a\w \iota_a *F - R*F + *(F_a\w R^a)) +c_3 R*F\big)\Big]. \label{inhomax-eq}
\ea
 
\section{PP-Wave Solutions}
We wish to explore solutions that characterize plane fronted waves with parallel rays. In Ehlers-Kundt form, we write the metric
\ba
g=2dudv + 2H(u,\zeta,\bar{\zeta})du^2 + 2d\zeta d\bar{\zeta},
\ea
where
\ba
u=\frac{z-t}{\sqrt{2}}, \quad v=\frac{z+t}{\sqrt{2}}, \quad \zeta=\frac{x+iy}{\sqrt{2}}
\ea
in terms of the Cartesian coordinates $(t,x,y,z)$.
$H$ is a smooth function which will be determined by Einstein field equations. We take next an electromagnetic potential 1-form given by $A=f(u,\zeta,\bar{\zeta})du$,
and the scalar field $\phi=\phi(u)$ only. 

\noindent
Introducing null tetrads to the formalism developed in Ref\cite{B11a} we have
\ba
l=du, \quad n=dv + Hdu, \quad m=d\zeta
\ea
The following notation used for the connection and curvature forms have been previously introduced in \cite{B16-a,B16-b}; then using the Cartan structure equations, we determine the connection 1-forms
\ba
\omega_+=0, \quad \omega_-=iH_{\zeta}l, \quad \omega_0=0,
\ea
the curvature 2-forms
\ba
R_+=0, \quad R_-=-iH_{\zeta\zeta}\; l\w m -iH_{\bar{\zeta}\bar{\zeta}} \;l \w \bar{m}, \quad R_0=0,
\ea
and the Einstein 3-forms
\ba
\frac{G_3+G_0}{\sqrt{2}}=2iH_{\zeta\bar{\zeta}} \; l\w m \w \bar{m}, \quad \frac{G_3-G_0}{\sqrt{2}}=0, \quad \frac{G_1+iG_2}{\sqrt{2}}=0.
\ea
Furthermore the electromagnetic field 2-form becomes
\ba
F=dA=-f_{\zeta}\; l \w m - f_{\bar{\zeta}}\; l \w \bar{m}.
\ea
Regardless of the choice of the non-minimal coupling constants $c_1, c_2$ and $c_3$, in this particular geometry the right hand sides of both (\ref{sc-eq}) and (\ref{inhomax-eq}) vanish. The scalar field equation is then identically satisfied $d(*d\phi)=0$.

\vspace{4mm}
\noindent

The Einstein equations after a long calculation reduce to
\ba
H_{\zeta\bar{\zeta}}&=&-\frac{f_{\zeta}f_{\bar{\zeta}}}{\phi} - \omega\frac{\phi_u^2}{2\phi^2}-\frac{\phi_{uu}}{2\phi}+ (c_1-c_2)\frac{f_{\zeta\zeta}f_{\bar{\zeta}\bar{\zeta}}}{\phi^2} +\frac{c_3}{\phi^2}f_{\zeta\bar{\zeta}}^2 \nonumber \\
&&+ \frac{(c_2-2c_3)}{\phi^2} (f_{\zeta\bar{\zeta}\bar{\zeta}}f_{\zeta}+2f_{\zeta\bar{\zeta}}^2+f_{\zeta\zeta\bar{\zeta}}f_{\bar{\zeta}}). \label{Ein-eq}
\ea

From the Maxwell equations $dF=0$ and $d*F=0$ we get
\ba
f_{\zeta\bar{\zeta}}=0.
\ea

Consequently the field equations reduce to
\ba
H_{\zeta\bar{\zeta}}=-\frac{f_{\zeta}f_{\bar{\zeta}}}{\phi}  -(\omega\frac{\phi_u^2}{2\phi^2}+\frac{\phi_{uu}}{2\phi})+ (c_1-c_2)\frac{f_{\zeta\zeta}f_{\bar{\zeta}\bar{\zeta}}}{\phi^2}. 
\ea

\vspace{2mm}
An exact solution of the of the above equation is
\ba
H(u,\zeta,\bar{\zeta})=H_0-\frac{1}{2\phi}f^2-(\omega\frac{\phi_u^2}{2\phi^2}+\frac{\phi_{uu}}{2\phi})\zeta\bar{\zeta}+(c_1-c_2)\frac{f_{\zeta}f_{\bar{\zeta}}}{\phi^2}
\ea
where $(H_0)_{\zeta\bar{\zeta}}=0$.
 \section{Conclusion}
 It is remarkable that for an arbitrary choice of the coupling constants $c_1, c_2$ and $c_3$, the right hand side of both the Maxwell and scalar field equations identically vanish in the pp-wave geometry. Then the scalar and vector fields $\phi$ and $F$ configurations of the non-minimal Brans-Dicke equations are still solutions. However, the Einstein field equations (\ref{Ein-eq}) are modified on its right hand side by non-minimal couplings. 

The previous case studied in \cite{B16} corresponds to the choice $c_1\neq 0$ and $c_2=c_3=0$ which give a non-trivial contribution in the metric function when integrated. We note the emergence of the scalar field coupling to the electromagnetic potential on the right hand side of the Einstein equations, which in turn leads to an exact solution that differs due to the conformal coupling. It may be interesting to include in future works a cosmological constant with a generalization of pp-waves in $AdS$ or $dS$. 

Here with the choice $c_1=c_2=\gamma$ and $c_3=\frac{\gamma}{3}$ we see that all non-minimal contributions drop out on-shell, i.e. provided the Maxwell equation $f_{\zeta\bar{\zeta}}=0$ is satisfied. This is a manifestation of the fact that the non-minimal coupling we study here is locally scale invariant.  The non-minimal couplings that we have considered here polarize spacetime and the presence of a dark matter distribution around a black hole will be affected for $c_1 \neq c_2$, which could provide dominant signs of the peculiar quality of dark matter\cite{B17}. Further studies\cite{B18} shows that the effects of cold dark matter on primordial gravitational waves; a frequency dependent modification, detectable but small, on the propagation speed. Consequently non-minimal couplings of $RF^2$ type may be of essence also in understanding the effects of magnetization and polarization on a binary inspiral. Remarkably, it has been pointed out recently that dark matter overdensities around black holes inescapably reshape the motion of binary mergers\cite{B19}.
 
\section{Acknowledgement}
Y.\c{S}. is grateful to Ko\c{c} University for its hospitality and partial support.




\begin{thebibliography}{99}

\bibitem{B1} C.H.Brans,R.H.Dicke,{\sl  Mach's principle and a relativistic theory of gravitation},Phys.Rev.{\bf 124},925(1961)   

\bibitem{B2} R.H.Dicke,{\sl Mach's principle and invariance under transformation of units},Phys.Rev.{\bf 125},2163 (1962) 

\bibitem{B3} C.H. Brans,{\sl Mach's principle and a relativistic theory of gravitation II},Phys.Rev.{\bf 125},2194 (1962)
\bibitem{B8}  T.Dereli,R.W.Tucker,{\sl Weyl scalings and spinor matter interactions in scalar-tensor theories of gravitation},Phys.Lett.{\bf B110}(1982)206


\bibitem{B7-a} A.R.Prassana {\sl } Phys.Lett.{\bf A37} (1971) 337
\bibitem{B7-b} G.W.Horndeski {\sl Conservation of charge and the Einstein-Maxwell field equations} J.Math.Phys. {\bf 17} (1976) 1980
\bibitem{B7-c} I.T.Dummond,S.J.Hathrell {\sl QED vacuum polarization in a background gravitational field and its effects on the velocity of photons} Phys.Rev.{\bf D22} (1980) 343
\bibitem{B7-d} A.B. Balakin,J.P.S.Lemos {\sl Non-minimal coupling for the electromagnetic fields: a general system of equations} Class.Quant.Gra. {\bf 22} (2005) 1867



\bibitem{B9-a} T.Dereli,Y.Senikoglu, {\sl Gravitational plane waves in a non-Riemannian description of Brans-Dicke gravity},Phys.Scr.{\bf 99}(2020) 045223

\bibitem{B9-b} M.Adak,T.Dereli,Y.Senikoglu, {\sl Non-Riemannian description of Robinson-Trautman spacetimes in Brans-Dicke theory of gravity},Int.J.Mod.Phys.{\bf D28}(2019) 1950070













\bibitem{B10} T.Dereli,Y.Senikoglu {\sl Screening of electric charges induced by non-minimal couplings of electromagnetic fields to gravity} (paper in progress)
\bibitem {B10b} T.Dereli,O.Sert, {\sl Nonminimal $ln(R)F^2$ couplings of electromagnetic fields to gravity: static,spherically symmetric solutions} Eur.Phys.J. {\bf C71} {2011} 1589

\bibitem{B11a} J.B.Griffiths,J.Podolsky,{\bf Exact Space-Times In Einstein's General Relativity}(Cambridge U.P.,2009),Chapter:17 




\bibitem {B16} T.Dereli,O.Sert, {\sl Nonminimally coupled gravitational and electromagnetic fields:pp-wave solutions} Phys.Rev. {\bf D83} {2011} 065005

\bibitem{B16-a} T.Dereli,R.W.Tucker, {\sl Exact neutrino solutions in the presence of torsion}, Phys.Lett.{\bf 82A}(1981) 229

\bibitem{B16-b} T.Dereli,A.A.Beler, {\sl Plane waves in supergravity}, Class.Quant.Grav. {\bf 2} (1985) 147-153 

\bibitem{B17} D.Psaltis, {\sl Two approaches to testing general relativity in the strong-field regime}, J. Phys. : Conf. Ser.,{\bf 189} (2009) 012033 
\bibitem{B18} R.Flauger,S.Weinberg, {\sl Gravitational Waves in Cold Dark Matter}, Phys.Rev.{\bf D97} (2018) 123506

\bibitem{B19} B.J.Kavanagh,D.A.Nichols,G.Bertone,D.Gaggero, {\sl Detecting dark matter around black holes with gravitational waves: Effects of dark-matter dynamics on the gravitational waveform},  arXiv {\bf2002.12811v1} [gr-qc]








\end{thebibliography}
\end{document}